# Observation of a ferro-rotational order coupled with second-order nonlinear optical fields

Wencan Jin[1,*], Elizabeth Drueke[1,*], Siwen Li[1], Alemayehu Admasu[2], Rachel Owen[1], Matthew Day[1], Kai Sun[1], Sang-Wook Cheong[2], Liuyan Zhao[1,+]

[1]*Department of Physics, University of Michigan, 450 Church St., Ann Arbor, Michigan, 48109*

[2]*Rutgers Center for Emergent Materials, Rutgers University, 136 Frelinghuysen Rd. Piscataway Township, New Jersey, 08854*

## Abstract

The ferro-rotational order [1-3], whose order parameter (OP) is an axial vector invariant under both time reversal (TR) and spatial inversion (SI) operations, is the last remaining category of ferroics to be observed after the ferroelectric, ferromagnetic, and ferro-toroidal orders. This order has become increasingly popular in many new quantum materials, especially in complex oxides [1,3], and is considered responsible for a number of novel phenomena such as polar vortices [4], giant magnetoelectric coupling [5], and type-II multiferroics [6]. However, physical properties of the ferro-rotational order have been rarely studied either theoretically or experimentally. Here, using high sensitivity rotational anisotropy second harmonic generation (RA SHG), we have, for the first time, exploited the electric quadrupole (EQ) contribution to the SHG to directly couple to this centrosymmetric ferro-rotational order in an archetype of type-II multiferroics, $RbFe(MoO_4)_2$. Surprisingly, we have found that two types of domains with opposite ferro-rotational vectors emerge with distinct populations at the critical temperature $T_c$ ~195 K and gradually evolve to reach an even ratio at lower temperatures. Moreover, we have identified the ferro-rotational order phase transition as weak first order, and have revealed its conjugate coupling field as a unique combination of the induced EQ SHG and the incident fundamental electric fields. Our results on physical properties of a ferro-rotational order provide crucial knowledge for understanding and searching for novel phases of matter built upon the ferro-rotational order. Further, these results open the possibility of revealing unconventional centrosymmetric orders and identifying their conjugate coupling fields with second order nonlinear optics.

The Landau theory of phase transitions [7,8], one major cornerstone in condensed matter physics research, introduces the concept of an OP, which attains a nonzero value below a phase transition, and classifies emergent phases by the symmetries under which they are invariant. An OP can be a scalar [9], a vector [3], or a higher-rank tensor [10] quantity that provides insight into the microscopic origin of its associated phase transition. Among all the symmetry operations, TR and SI symmetries are of particular importance. They categorize OPs into four groups based on even (+) or odd (−) parity under TR and SI operations [3,11,12], and this classification guides the search for new phases and their conjugate coupling fields [1-3,13].

Ferroics with vector OPs are a class of materials of particular interest, as two of the most well-known ordered phases, ferroelectricity and ferromagnetism, belong to this family and couple to the two most common fields, the electric ($\vec{E}$) and magnetic ($\vec{B}$) fields, respectively. Within the categorization framework based on OP parities under TR and SI operations, electric polarization ($\vec{P}$) and magnetization ($\vec{M}$), OPs for ferroelectricity and ferromagnetism, are TR+SI− (*i.e.*, + −) and − +, respectively, while the toroidal arrangements of electric ($\vec{r} \times \vec{P}$) [3] and magnetic ($\vec{r} \times \vec{M}$) [11,14] dipoles, OPs for ferro-rotational and ferro-toroidal orders, are + + and − −, respectively, as outlined in Fig. 1(a). In contrast to ferroelectricity and ferromagnetism, the search for ferro-rotational and ferro-toroidal orders has been challenging because they are much rarer and the conjugate fields to their OPs are not readily available. In fact, it was not until very recently that the ferro-toroidal order was detected by optical SHG by exploiting its broken SI symmetry[11] and its conjugate field was determined to be $\vec{E} \times \vec{B}$ with hysteretic poling behaviors [12]. Meanwhile, the ferro-rotational order is known to be present in complex oxides with structural distortions of uniform oxygen cage rotations [1,3,5,15-17], but its symmetry properties, domain structures, and conjugate field have remained elusive.

SHG is a process in which light frequency is doubled through its second order nonlinear interactions with a material. It has so far been widely used to detect phases with broken SI symmetry only in which the leading-order electric dipole (ED) contribution to SHG is allowed [11,18-23]. Recently, the development of high sensitivity SHG makes it possible to detect SHG from higher-order multipolar contributions, such as EQ, in SI symmetry-preserved states [24-27]. However, to our best knowledge, it has not yet been shown to reveal a phase transition of a centrosymmetric order, such as the ferro-rotational order, using EQ SHG. (see EQ SHG measurement details in Methods).

As depicted in Fig. 1(b), RbFe(MoO$_4$)$_2$, an archetype of type-II multiferroic materials predicted to host a ferro-rotational order [6,15,16,28,29], consists of stacks of FeO$_6$ octahedra sharing vertices with MoO$_4$ tetrahedra. It undergoes a structural phase transition at $T_c$ ~195 K with octahedra (tetrahedra) rotating counterclockwise (clockwise) about the $c$ axis for domain type I and vice versa for domain type II. The point group of the room temperature phase of RbFe(MoO$_4$)$_2$ is known to be $\bar{3}m$, while that for the low temperature structural phase is likely to be $\bar{3}$ but with a couple of other options including $3m$ and $32$ [16,28,30] (see sample growth details in Methods). Furthermore, the physical properties of this low temperature structural phase have been hardly explored.

We first establish the EQ contribution to the SHG from the centrosymmetric phase of RbFe(MoO$_4$)$_2$ at room temperature. In order to access all nonzero second order susceptibility tensor elements, we performed RA SHG measurements $I^{2\omega}(\phi)$ at room temperature in all four polarization combination channels, namely P/S$_{in}$ − P/S$_{out}$, in the oblique incidence geometry (Fig. 2(b)), where P/S$_{in/out}$ stands for the polarization of incoming/reflected (in/out) light being parallel/normal (P/S) to the light scattering plane. Given that the room temperature bulk point group $\bar{3}m$ contains SI symmetry, there can be no leading-order ED contribution to $I^{2\omega}(\phi)$, and the next highest order contribution, EQ, must be considered, which is

$$P_i^{eff}(2\omega) = \chi_{ijkl}^{EQ} E_j(\omega) \partial_k E_l(\omega) \tag{1}$$

Qualitatively, the four patterns in Fig. 2(a) all exhibit a three-fold rotational (C$_3$) symmetry and three mirror planes at 90°, 210° and 330°, which is consistent with $\bar{3}m$. Quantitatively, the RA SHG patterns can be well fitted with the $I^{2\omega}(\phi)$ functions derived from the EQ SHG process under $\bar{3}m$, in which

$$I^{2\omega}_{P/S_{in}-P_{out}}(\phi) \propto (C_1 + C_2 \sin(3\phi))^2 + (C_3 + C_4 \sin(3\phi))^2$$
$$I^{2\omega}_{P/S_{in}-S_{out}}(\phi) \propto (C_5 \cos(3\phi))^2 \tag{2}$$

with $C_{1,2,3,4,5}$ being linear combinations of $\chi_{ijkl}^{EQ}$, the EQ SHG susceptibility tensor elements (see Supplementary Section 1). We further rule out the possibilities of surface ED SHG and electric field induced SHG (EFISH) (see Supplementary Section 2), and therefore identify the bulk EQ SHG under centrosymmetric $\bar{3}m$ as being responsible for the SHG measured at room temperature.

Having determined the EQ origin of the SHG from RbFe(MoO$_4$)$_2$ at room temperature, we proceed to perform RA SHG measurements in the normal incidence geometry to select a subset of $\chi^{EQ}_{ijkl}$ tensor elements. Figure 2(b) shows two RA SHG patterns taken with incoming and reflected light polarizations in parallel and crossed configurations at normal incidence. Like the oblique incidence results, both patterns possess the C$_3$ axis and the three mirror planes which are enforced by the bulk point group $\bar{3}m$. In addition, the simulated functions for the normal incidence geometry are significantly simplified from those of the oblique incidence geometry to

$$I^{2\omega}_{parallel,high}(\phi) = (\chi^{EQ}_{yyzy} \cos(3\phi))^2$$
$$I^{2\omega}_{cross,high}(\phi) = (\chi^{EQ}_{yyzy} \sin(3\phi))^2 \qquad (3)$$

where only one tensor element $\chi^{EQ}_{yyzy}$ is selected out. This allows us to uniquely extract the value of $\chi^{EQ}_{yyzy}$ and track its evolution with varying temperatures, which motivates us to use the normal incidence geometry to investigate the temperature dependence.

Figure 3(a) shows the RA SHG patterns taken in the parallel channel of the normal incidence geometry at selected temperatures decreasing from 200 K to 80 K. The evolution of these patterns clearly demonstrates a phase transition occurring at a $T_c$ between 200 K and 190 K, as evidenced by both the sudden appearance of a nonzero background in the RA SHG patterns and the start of the rotation of the RA SHG patterns away from the mirror planes of the high temperature phase.

We first identify unequivocally the point group of the RbFe(MoO$_4$)$_2$ low temperature phase, which until now has been debated because it is challenging for infrared spectroscopy and x-ray diffraction [30] to distinguish the subtle differences between point groups $3m$, $32$, and $\bar{3}$. The $3m$ point group is immediately disqualified because the departure of the RA patterns from the mirror planes at room temperature indicates the absence of mirror symmetries below $T_c$. The non-centrosymmetric $32$ point group can also be reliably ruled out, because the SHG intensity remains nearly unchanged across $T_c$ instead of increasing by orders of magnitude as it would be expected from the ED contribution allowed in a system with broken SI symmetry. A comparison of the simulated RA SHG patterns for $3m$ and $32$ (Fig. 3(c)) with our experimental observation further serves to rule out these two point groups, as the simulated results show perfect alignments with the three mirror planes while the experimental result does not. Thus, the only possible point group left for the low temperature phase of RbFe(MoO$_4$)$_2$ is $\bar{3}$, which breaks the three mirrors while preserving the SI and C$_3$ symmetries. The bulk EQ SHG under $\bar{3}$ scales as

$$I^{2\omega}_{parallel,D1}(\phi) = (\chi^{EQ}_{xxzx} \sin(3\phi) + \chi^{EQ}_{yyzy} \cos(3\phi))^2 \qquad (4)$$

which accounts for the rotation, but not the nonzero background of the RA SHG patterns below $T_c$.

We next prove the presence of two types of domains in RbFe(MoO$_4$)$_2$ at low temperatures by analyzing the nonzero background in the RA SHG patterns that appears below $T_c$. As the $I^{2\omega}_{parallel,D1}(\phi)$ pattern rotates counterclockwise by an angle

$$\delta = \frac{1}{3}\tan^{-1}\left(\frac{\chi^{EQ}_{xxzx}}{\chi^{EQ}_{yyzy}}\right) \tag{5}$$

its broken-mirror-related counterpart is expected to rotate clockwise by the same angle $\delta$, yielding a functional form of

$$I^{2\omega}_{parallel,D2}(\phi) = (-\chi^{EQ}_{xxzx}\sin(3\phi) + \chi^{EQ}_{yyzy}\cos(3\phi))^2 \tag{6}$$

These two functions correspond to the two types of domains with opposite ferro-rotational vectors, where the FeO$_6$ octahedra rotate counterclockwise and clockwise, respectively, as depicted in Fig. 1(b) for $T < T_c$. The nonzero background, which cannot be accounted for by either individual type of domains, can be well explained by a weighted linear superposition of contributions from both types of domains,

$$I^{2\omega}_{parallel,low}(\phi) = A \cdot I^{2\omega}_{parallel,D1}(\phi) + (1-A) \cdot I^{2\omega}_{parallel,D2}(\phi) \tag{7}$$

where $A$ is the weight of one type of domains. Figure 3(b) shows one example of the RA SHG pattern at 170 K well fitted by this weighted domain averaged model. The data indicates a ferro-rotational domain size much smaller than our SHG probe beam diameter of 50 $\mu$m.

We now are ready to explore the temperature dependence of $\chi^{EQ}_{xxzx}$, $\chi^{EQ}_{yyzy}$, domain weight $A$, and the RA SHG pattern rotation angle $\delta$ with best fits of the domain averaged model to the RA SHG patterns at every temperature measured between 80 K and 210 K to the domain averaged model. As shown in Fig. 4(a), $\chi^{EQ}_{xxzx}$ jumps from zero to a finite value at $T_c$ ~195 K and gradually grows larger before saturating at lower temperatures, suggesting this structural phase transition is of weak first order character and $\chi^{EQ}_{xxzx}$ is, to the lowest order approximation, linearly proportional to the ferro-rotational OP. In contrast, $\chi^{EQ}_{yyzy}$ in Fig. 4(b) is present above $T_c$, experiences a sharp spike at $T_c$, and decreases slowly below $T_c$. The temperature dependence of domain populations in Fig. 4(c) suggests that the two types of domains show up with uneven populations at $T_c$ and converge to equal populations with decreasing temperature. Lastly, the RA SHG rotation angle $\delta$, shown in Fig. 4(d), exhibits a jump from 0° to 10° at $T_c$ and gradually approaches its maximum of 20° at lower temperatures, mimicking $\chi_{xxzx}^{EQ}$.

At this point, we have used EQ SHG for the first time to reveal a ferro-rotational order in RbFe(MoO$_4$)$_2$, by unambiguously identified its centrosymmetric point group $\bar{3}$, revealing its two types of domains, and showing its temperature dependences. In the following, we will use phenomenological Landau theory to first explain the coupling between the EQ SHG fields and this ++ ferro-rotational OP, and then to understand the temperature dependence of the EQ SHG susceptibility tensor elements. To begin, because the point group $\bar{3}$ of the ferro-rotational order is a subgroup of $\bar{3}m$, its OP therefore transforms as the $A_{2g}$ symmetry of $\bar{3}m$. Under the plane wave approximation, the radiated EQ SHG fields in the normal incidence geometry are expressed as:

$$\begin{pmatrix} E_x^{2\omega} \\ E_y^{2\omega} \end{pmatrix} \propto \begin{pmatrix} P_{x,eff}^{2\omega} \\ P_{y,eff}^{2\omega} \end{pmatrix} = \chi_{xxzx}^{EQ} \begin{pmatrix} (E_x^\omega)^2 - (E_y^\omega)^2 \\ -2 E_x^\omega E_y^\omega \end{pmatrix} k_z + \chi_{yyzy}^{EQ} \begin{pmatrix} 2 E_x^\omega E_y^\omega \\ (E_x^\omega)^2 - (E_y^\omega)^2 \end{pmatrix} k_z \quad (8)$$

where $E_{x/y}(\omega)$ ($E_{x/y}(2\omega)$) is the incident fundamental (radiated SHG) electric fields, and $k_z$ is the wave vector of the incident electric fields. From this, we find that $\chi_{xxzx}^{EQ}$ and $k_z \left( E_x^{2\omega}(E_x^\omega)^2 - E_x^{2\omega}(E_y^\omega)^2 - 2 E_x^\omega E_y^\omega E_y^{2\omega} \right)$ share the same symmetry, both transforming as the $A_{2g}$ symmetry of $\bar{3}m$. It now becomes clear that a proper combination of the induced EQ SHG fields and the incident fundamental electric fields, such as $k_z \left( E_x^{2\omega}(E_x^\omega)^2 - E_x^{2\omega}(E_y^\omega)^2 - 2 E_x^\omega E_y^\omega E_y^{2\omega} \right)$ above, has the same symmetry as the ferro-rotational order OP, and therefore forms a conjugate coupling field to the OP. It is worth noting that this is the lowest order combination of the polar vector field combination to achieve the same symmetry as the ferro-rotational OP, and it at least requires the EQ SHG process to observe the ferro-rotational order in RbFe(MoO$_4$)$_2$.

In order to capture the temperature dependence of the EQ SHG susceptibility tensor elements, we start by expanding the Landau free energy in terms of the ferro-rotational order OP, $\eta$, as

$$F(T) = F_0(T) + \alpha(T - T_C)\eta^2 + \beta\eta^4 + \gamma\eta^6 \quad (9)$$

where $\alpha > 0$, $\beta < 0$, and $\gamma > 0$ are constants near $T_c$ for the small $\eta$ near the weak first order phase transition [8,31]. Minimizing this free energy yields a functional form for the temperature dependence of $\eta$ of

$$\eta(T) = \begin{cases} 0, & T > T_c \\ \sqrt{a + b\sqrt{T_d - T}}, & T \leq T_c \end{cases}$$

(10)

where $a = -\frac{\beta}{3\gamma}$, $b = \sqrt{\frac{\alpha}{3\gamma}}$ and $T_d = \frac{\beta^2}{3\alpha\gamma} + T_c$ [31]. From Equation (8), we learn that $\chi^{EQ}_{xxzx}$, just like $\eta$, obeys the $A_{2g}$ symmetry of $\bar{3}m$, while $\chi^{EQ}_{yyzy}$ is of the $A_{1g}$ symmetry. We therefore can expand the two tensor elements as

$$\chi^{EQ}_{xxzx} = a_1\eta + a_3\eta^3 + \cdots \quad (11)$$

$$\chi^{EQ}_{yyzy} = a_0 + a_2\eta^2 + a_4\eta^4 \ldots \quad (12)$$

and arrive at their expected temperature dependences with which the raw data in Fig. 4(a) and (b) are well fit. These fits give $T_c = 194.5 \pm 0.9$ K and $T_d = 199.6 \pm 2.1$ K. Subsequently, the temperature dependence of $\delta$ in Fig. 4(d) is also nicely explained.

In conclusion, we have here presented our symmetry-resolved and temperature-dependent SHG study on a TR+SI+ ferro-rotational order in RbFe(MoO$_4$)$_2$. Unlike conventional SHG which relies on the ED contribution from broken SI symmetry, we have exploited the EQ SHG process to monitor this SI symmetry-preserved phase transition and unequivocally determined the symmetry point group $\bar{3}$ for this centrosymmetric ferro-rotational ordered phase below $T_c$ in RbFe(MoO$_4$)$_2$. We have successfully revealed the presence of two types of ferro-rotational order domains with opposite OP vectors and shown their uneven population right below $T_c$ evolving to even at lower temperature. Furthermore, we identified the conjugate coupling field, a proper combination of the induced EQ SHG and the incident fundamental electric fields, for this $++$ axial vector OP, and tracked the temperature dependence of the OP through the EQ SHG susceptibility tensor elements and the domain populations through the domain averaged model. The identification of this last remaining vector-type OP in ferroic orders its conjugate coupling fields is of crucial importance for understanding and searching for novel phases built upon ferro-rotational orders, such as the type-II multiferroic order developed in RbFe(MoO$_4$)$_2$ at even lower temperature [15,29,32]. Moreover, our work also introduces the possibility of using second order nonlinear optical measurements to probe inversion symmetric novel phases of matter, and showcases an example of evaluating conjugate fields for unconventional OPs whose coupling fields are not easily identifiable.

**Methods**

**Growth of RbFe(MoO$_4$)$_2$ single crystals.** Single crystals of RbFe(MoO$_4$)$_2$ were synthesized by spontaneous crystallization from the flux melt method as described in Ref. [16]. Powders of (Alfa Aesar, 5N purity) Rb$_2$CO$_3$, Fe$_2$O$_3$ and MoO$_3$ were thoroughly mixed in the molar ratio of 2:1:6. The homogenized mixture was heated in a platinum crucible at 1100 K for 20 h in air and was cooled at a rate of 2K/h to 900 K followed by subsequent faster cooling at 5K/h to room temperature. Transparent light-yellow to light-green hexagonal platelet crystals of typical dimensions of 3×3×0.1 mm$^3$ were readily separated from the flux for experiments by dissolving in warm water.

**RA SHG measurements.** In the RA SHG measurement with the oblique (normal) incidence geometry, the reflected SHG intensity is recorded as a function of the azimuthal angle $\phi$ between the scattering plane (electric polarization) and the in-plane crystalline axis. In this experiment, the incident ultrafast light source was of 800 nm wavelength, 40 fs pulse duration and 200 kHz repetition rate, and was focused onto a 50 $\mu$m diameter spot on the sample with a fluence of ~ 0.25 mJ/cm$^2$. The intensity of the reflected SHG was measured with a single photon counting detector. All thermal cycles were carried out with a base pressure better than 5×10$^{-7}$ mbar.

**Data availability**

The datasets generated and/or analyzed during the current study are available from the corresponding author on reasonable request.

[*] authors contributed equally

[+] corresponding to: lyzhao@umich.edu

**Acknowledgements**

We acknowledge technical assistance from K. Mattioli. L. Zhao acknowledges support by NSF CAREER Grant No. DMR-174774. E. Drueke acknowledges support by NSF Graduate Research Fellowship Program under Grant No. DGE-1256260. S. Cheong acknowledges that the work at Rutgers is funded by the Gordon and Betty Moore Foundation's EPiQS Initiative through Grant GBMF4413 to the Rutgers Center for Emergent Materials. K. Sun acknowledges support by NSF Grant No. NSF-EFMA-1741618 and the Alfred P. Sloan Foundation.

**Author contributions**

W. J., S.-W. C., and L. Z. conceived and initiated this project; A. A. and S.-W. C. synthesized the bulk RbFe(MoO$_4$)$_2$ crystals; W. J., E. D., and S. L. performed RA SHG measurements; W. J. and E. D. carried out the Landau theory analysis under the guidance of K. S. and L. Z.; W. J., E. D., and L. Z. analyzed the data and wrote the manuscript; all authors participated in the discussion of the results.

**Competing interests**

The authors declare no competing interests.

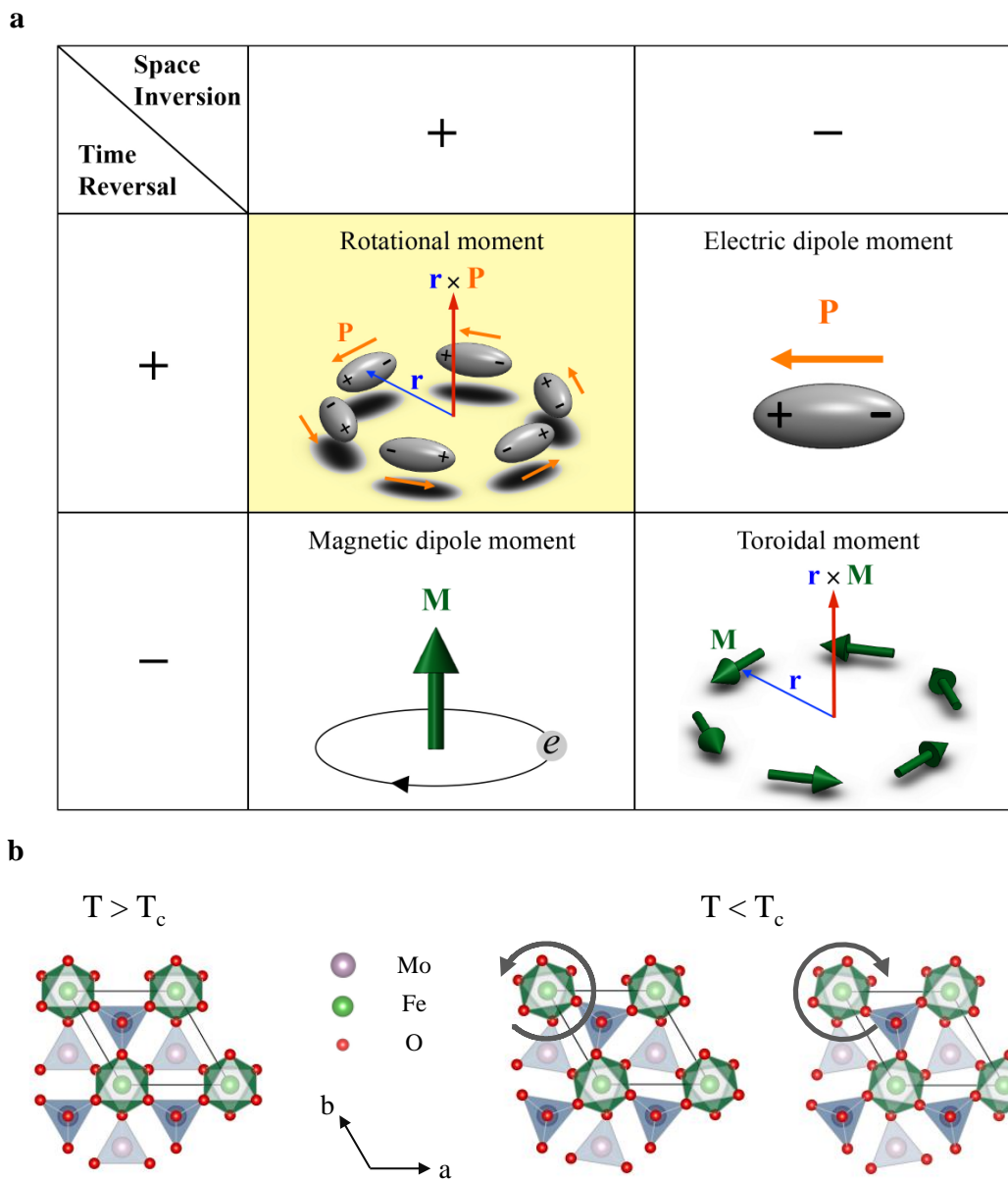

**Figure 1. Categorizing ferroic orders with vector order parameters. a**, A summary of the four vector OPs classified by their parities under TR and SI operations. Here + indicates even parity and − indicates odd parity. The yellow background highlights the ferro-rotational OP. **b**, The crystal structure of $RbFe(MoO_4)_2$ as viewed along the $c$ axis, both above and below the structural phase transition temperature $T_c$. Two types of domains are expected below $T_c$, corresponding to counterclockwise and clockwise rotations of the $FeO_6$ octahedra.

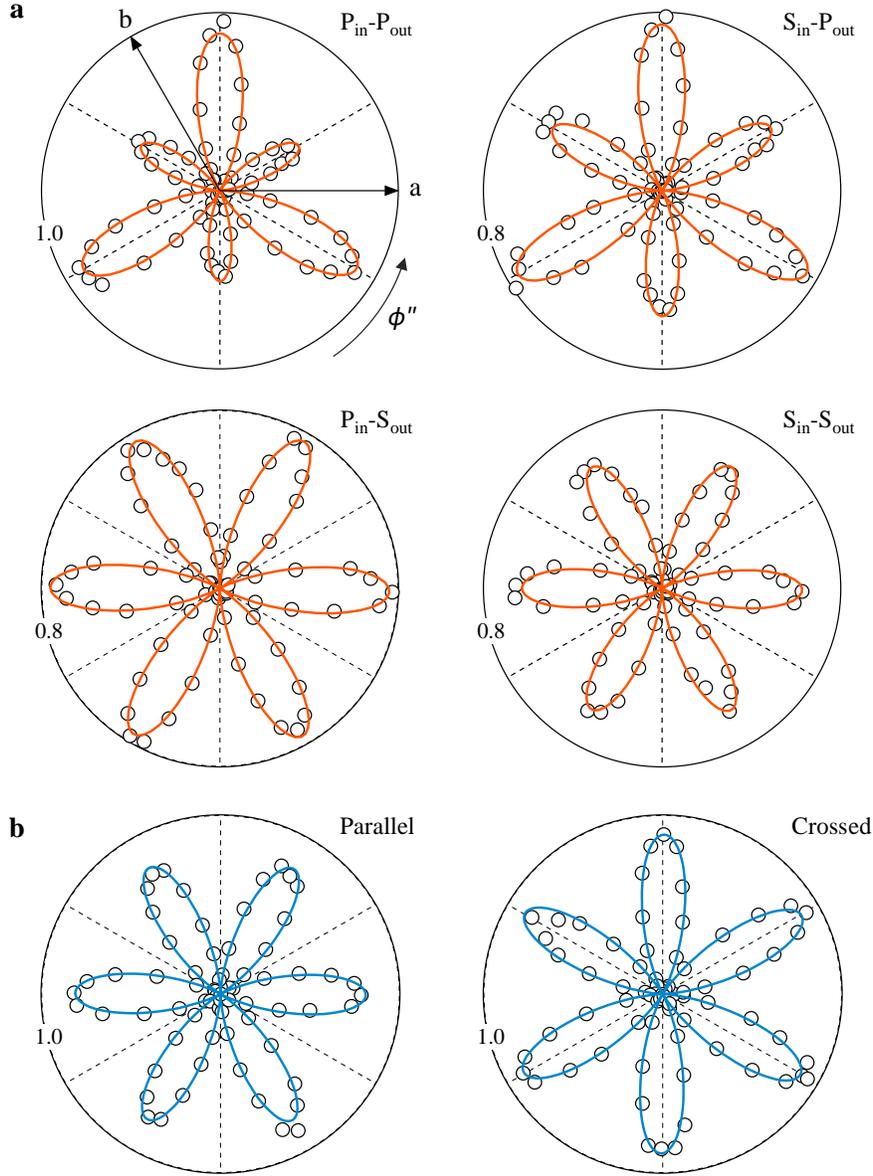

**Figure 2. Identifying the bulk EQ contribution to the room temperature SHG.** Polar plots of room temperature ($T$ = 290 K) RA SHG patterns fitted with the functional forms derived from the bulk EQ SHG susceptibility tensor under point group $\bar{3}m$, **a**, at oblique incidence in all four polarization combination channels, namely P/$S_{in}$ – P/$S_{out}$, and **b**, at normal incidence in both the parallel and crossed polarization channels. Open circles indicate the raw RA SHG data and the solid curves are for the bulk EQ SHG fits. The crystalline axes *a*, *b* are defined in Fig. 1b, labeled here in $P_{in}$ – $P_{out}$ and omitted for the rest. The three vertical mirror planes in $\bar{3}m$ are indicated by the three dashed radial lines in every plot. All data sets are plotted on the same intensity scale, normalized to a value of 1 corresponding to 22 fW.

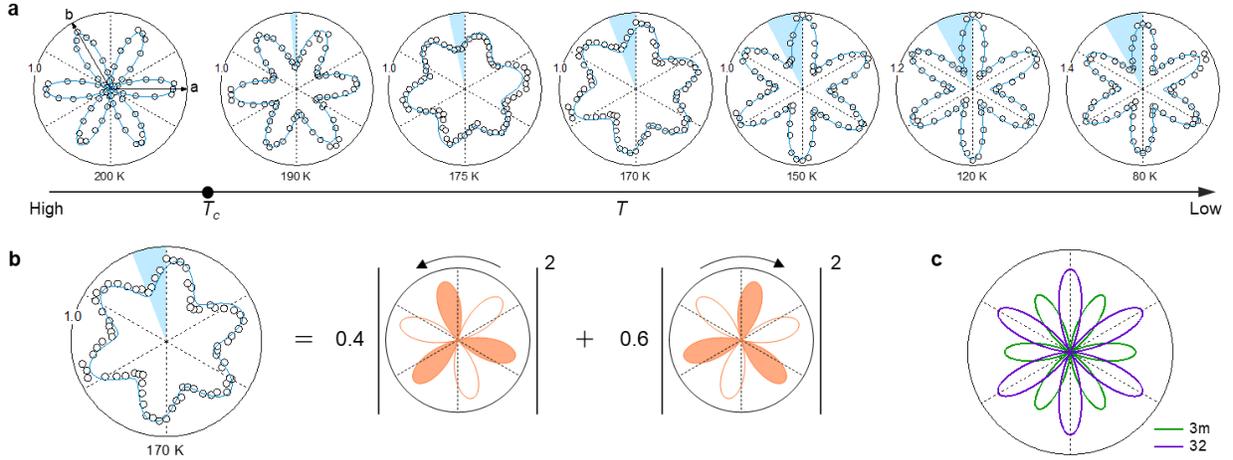

**Figure 3. Tracking the temperature dependence of the ferro-rotational order. a**, Polar plots of the RA SHG data in the parallel channel of the normal incidence geometry at selected temperatures above and below $T_c$. The rotation of each pattern away from the room temperature vertical mirror at 90° is highlighted by blue shading. The pattern above $T_c$ is fitted to the bulk EQ SHG under $\bar{3}m$, and the patterns below $T_c$ are fitted to a weighted two-domain averaged model of the bulk EQ SHG under $\bar{3}$. All data sets are plotted on the same intensity scale, normalized to a value of 1 corresponding to 22 fW. **b**, An example of fitting the RA SHG pattern at 170 K, below $T_c$, with a weighted average of contributions from both types of domains in Fig. 1b. The individual patterns (orange) from the two domains rotate counterclockwise and clockwise, respectively, and their contributions to the total RA SHG are indicated as the coefficients, 0.4 and 0.6, respectively. The solid and empty petals represent the phases for the SHG electric fields. **c**, Polar plots of the simulated RA SHG patterns under point groups $3m$ (green) and $32$ (purple).

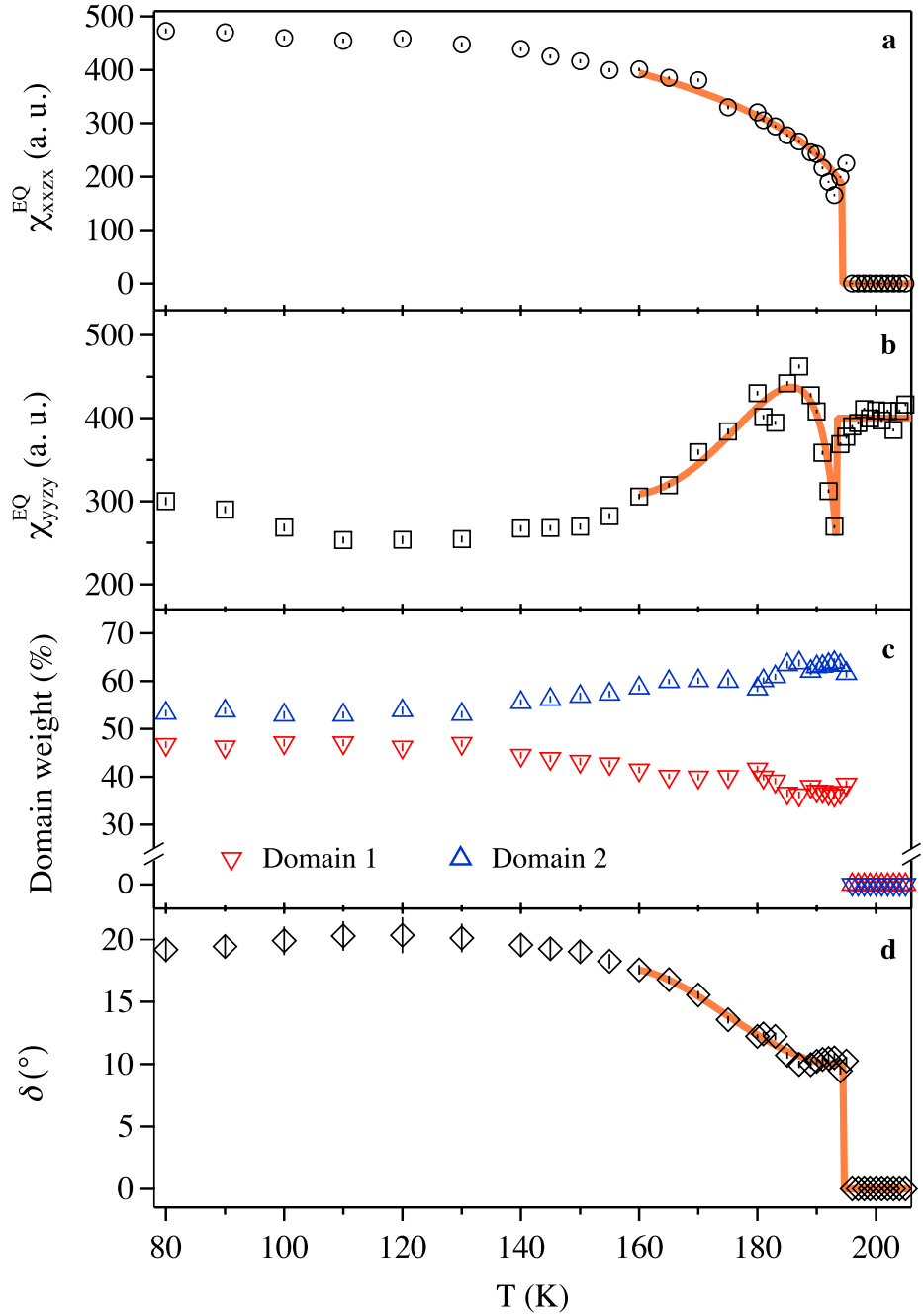

**Figure 4. Resolving the temperature dependence of fitting parameters.** Temperature dependence of EQ SHG susceptibility tensor elements **a**, $\chi^{EQ}_{xxzx}$ and **b**, $\chi^{EQ}_{yyzy}$, **c**, the weight of both domains, and **d**, the rotation of the RA SHG pattern for a single domain, $\delta$. The open circle (a), square (b), triangle (c), and diamond (d) are unique fit values from the RA SHG data at varying temperatures. The orange solid lines in (a), (b), and (d) are best fit to the Landau theory based functional forms for these parameters. The error bars stand for one standard error in fitting the RA SHG data with the domain averaged model.

# Supplementary Information
# Observation of a ferro-rotational order coupled with second-order nonlinear optical fields

Wencan Jin[1,*], Elizabeth Drueke[1,*], Siwen Li[1], Alemayehu Admasu[2], Rachel Owen[1], Matthew Day[1], Kai Sun[1], Sang-Wook Cheong[2], Liuyan Zhao[1,+]

[1]Department of Physics, University of Michigan, 450 Church St., Ann Arbor, Michigan, 48109

[2]Rutgers Center for Emergent Materials, Rutgers University, 136 Frelinghuysen Rd. Piscataway Township, New Jersey, 08854

**Table of Contents**

S1.     Simulations of the bulk EQ SHG under $\bar{3}m$ in the oblique incidence geometry

S2.     Simulations of the electric field induced SHG under $\bar{3}m$ in the oblique incidence geometry

[*] authors contributed equally

[+] corresponding to: lyzhao@umich.edu

## S1. Simulations of the bulk EQ SHG under $\bar{3}m$ in the oblique incidence geometry

We have simulated the functional forms for the bulk EQ SHG under the point group $\bar{3}m$ in the oblique incidence geometry, using

$$I^{2\omega}(\phi) = \left|A\hat{e}_i(2\omega)\chi_{ijkl}^{EQ}(\phi)\hat{e}_j(\omega)\hat{\partial}_k(\omega)\hat{e}_l(\omega)\right|^2 I^\omega I^\omega \tag{S1}$$

where $A$ is a constant determined by the experimental geometry, $I^\omega$ is the intensity of the incident beam, $\hat{e}$ is the polarization of the incoming fundamental or outgoing SHG light, and $\chi_{ijkl}^{EQ}(\phi)$ is the bulk EQ susceptibility tensor transformed into the rotated frame of the scattering plane. The nonzero independent elements of the tensor in the unrotated frame of the crystal are deduced by applying the point group $\bar{3}m$ and degenerate SHG permutation symmetries. This reduces $\chi_{ijkl}^{EQ}(\phi = 0)$ to 11 nonzero independent elements

$$xxxx = yyyy = yyxx + yxxy + yxyx;$$
$$yyxx = xxyy = yxxy = xyyx; \; yxyx = xyxy;$$
$$xxzz = yyzz = xzzx = yzzy; \; zzxx = zzyy = zxxz = zyyz;$$
$$yyyz = -yxxz = -xyxz = -xxyz = yzyy = -yzxx = -xzyx = -xzxy;$$
$$yyzy = -yxzx = -xyzx = -xxzy; \; zyyy = -zyxx = -zxyx = -zxxy;$$
$$xzxz = yzyz; zxzx = zyzy;$$
$$zzzz; \tag{S2}$$

Combining Equations (S1) and (S2), we obtain the functional forms for all four polarization combination channels as follows, taking $I^\omega$ and $\hat{\partial}_k$ as unities:

$$I_{P_{in}-P_{out}}^{2\omega}(\phi) = \cos^2\theta\left\{\sin^2\theta\left[\chi_{zyzy}\cos^2\theta + (-2\chi_{zzyy} + \chi_{zzzz})\sin^2\theta\right.\right.$$
$$\left.+ \chi_{zyyy}\cos\theta\sin\theta\sin(3\phi)\right]^2$$
$$+ \left[(2\chi_{xxyy} + \chi_{xyxy} - 2\chi_{yyzz})\cos^2\theta\sin\theta + \chi_{yzyz}\sin^3\theta\right.$$
$$\left.\left.+ (\chi_{yyzy}\cos^3\theta - 2\chi_{yyyz}\cos\theta\sin^2\theta)\sin(3\phi)\right]^2\right\}$$

$$\tag{S3}$$

$$I^{2\omega}_{S_{in}-P_{out}}(\phi) = \cos^2\theta[\chi_{xyxy}\sin\theta - \chi_{yyzy}\cos\theta\sin(3\phi)]^2$$
$$+ \sin^2\theta[\chi_{zyzy}\cos\theta - \chi_{zyyy}\sin\theta\sin(3\phi)]^2$$

(S4)

$$I^{2\omega}_{P_{in}-S_{out}}(\phi) = [(\chi_{yyzy}\cos^3\theta - 2\chi_{yyyz}\cos\theta\sin^2\theta)\cos(3\phi)]^2$$

(S5)

$$I^{2\omega}_{S_{in}-S_{out}}(\phi) = [\chi_{yyzy}\cos\theta\cos(3\phi)]^2$$

(S6)

where $\theta$ is the incident angle ($\theta = 0$ for normal incidence). Equations (S3-6) are then short written into the forms in Equation (2) of the main text.

## S2. Simulations of the electric field induced SHG under $\bar{3}m$ in the oblique incidence geometry

We have simulated the functional forms for the electric field induced SHG (EFISH) under $\bar{3}m$ in the oblique incidence geometry, using

$$I^{2\omega}(\phi) = \left| A\hat{e}_i(2\omega)\chi_{ijkl}^{EQ}(\phi)\hat{e}_j(\omega)\vec{E}_{k=z}\hat{e}_l(\omega) \right|^2 I^\omega I^\omega \tag{S7}$$

Comparing to Equation (S1), we see here that the gradient along $k = x, y, z$ component has been replaced by the DC electric field $\vec{E}$ that is normal to the sample surface, i.e., $k = z$.

Using Equations (S2) and (S7), we obtain the functional forms for the EFISH in all four polarization combination channels as follows, taking $I^\omega$ and $\vec{E}_{k=z}$ as unities:

$$I^{2\omega}_{P_{in}-P_{out}}(\phi) = \left[\chi_{zyzy}\cos^2\theta\sin\theta + \chi_{zzzz}\sin^3\theta\right]^2$$
$$+ \left[\cos^2\theta(-2\chi_{yyzz}\sin\theta + \chi_{yyzy}\cos\theta\sin(3\phi))\right]^2 \tag{S8}$$

$$I^{2\omega}_{S_{in}-P_{out}}(\phi) = (\chi_{zyzy}\sin\theta)^2 + (\chi_{yyzy}\cos\theta\sin(3\phi))^2 \tag{S9}$$

$$I^{2\omega}_{P_{in}-S_{out}}(\phi) = \left[\chi_{yyzy}\cos^2\theta\cos(3\phi)\right]^2 \tag{S10}$$

$$I^{2\omega}_{S_{in}-S_{out}}(\phi) = \left[\chi_{yyzy}\cos(3\phi)\right]^2 \tag{S11}$$

where $\theta$ is the same incident angle defined in Equations (S3-6). In contrast to Equation (S4) for the $S_{in} - P_{out}$ EQ SHG, which has alternating SHG lobes in its polar plot, Equation (S9) expects six even lobes in the $S_{in} - P_{out}$ EFISH polar plot. In Figure S1, the EQ SHG model clearly fits the experimental data better than the EFISH model for the $S_{in} - P_{out}$ RA SHG at room temperature due to these alternating peak intensities.

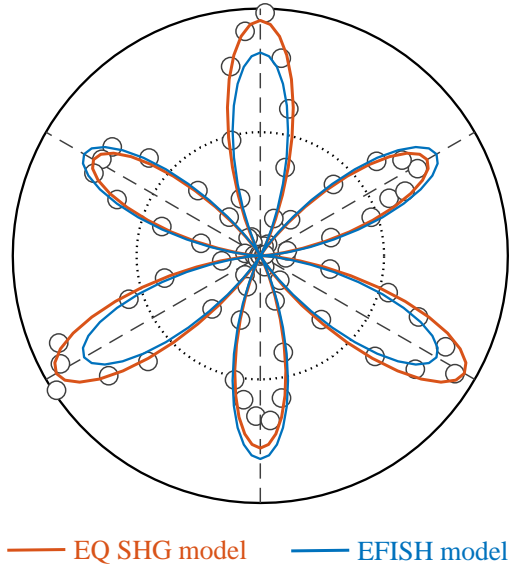

**Figure S1. Bulk EQ SHG vs EFISH.** Fitting the room temperature RA SHG data in the $S_{in} - P_{out}$ channel using the bulk EQ SHG (red) and EFISH (blue) models.